\documentstyle[12pt]{article}
\newcommand{\tline}{\setlength{\baselineskip}{0.84cm}}

\begin{document}
\tline
\tline
\setcounter{page}{1}
\begin {center}
{\large {\bf On the Proof by {\it Reductio ad Absurdum}}}\\
{\large {\bf of the Hohenberg-Kohn Theorem for Ensembles of Fractionally Occupied States of Coulomb Systems}} \\
\end{center}
\vspace{0.75cm}
\begin{center} 
Eugene S. KRYACHKO\footnote[1]{Address for correspondence: FAX: +32 (4) 366 3413; E-mail address: eugene.kryachko@ulg.ac.be}
\end{center}
\vspace{0.5cm}
{\normalsize \centerline{Bogoliubov Institute for Theoretical Physics, Kiev, 03143 Ukraine}}
{\normalsize \centerline{and}}
{\normalsize \centerline{Department of Chemistry, Bat. B6c, University of Liege}}
{\normalsize \centerline{Sart-Tilman, B-4000 Liege 1, Belgium}}
\vspace{1cm}
\begin{abstract}
\vspace{0.25cm}It is demonstrated that the original {\em reductio ad absurdum} proof of the generalization of the Hohenberg-Kohn theorem for ensembles of fractionally occupied states for isolated many-electron Coulomb systems with Coulomb-type external potentials 
by Gross et al. [Phys. Rev. A {\bf 37}, 2809 (1988)] is self-contradictory since the to-be-refuted assumption (negation) regarding the ensemble one-electron densities and the assumption about the external potentials are logically incompatible to each other due to the Kato electron-nuclear cusp theorem. It is however proved that the Kato theorem itself provides a satisfactory proof of this theorem. 
\end{abstract}
\vspace{1cm}
\vspace{1cm}

\newpage 

The Hohenberg-Kohn theorem [1] that underlies the foundation of the density functional theory (see Ref. [2] and references therein) has been generalized in a number of ways, particularly for ensembles of fractionally occupied states by Gross et al. [3] (EGHK theorem) and for the degenerate ground states by van Leeuwen [4] (DGHK theorem). Both these theorems have been proved invoking the {\em reductio ad absurdum} method that was used in the original proof of the Hohenberg-Kohn theorem [1,5].   

By analogy with the recent work [5], the present Note re-examines the proof by {\em reductio ad absurdum} of the EGHK theorem and demonstrates that, although its statement is generally correct, its original proof cannot be maintained if the external potential is of Coulomb type because otherwise it implies that the supposed ensemble one-electron densities should violate the Kato electron-nuclear cusp conditions provided by the Kato theorem. It is however proved that the Kato theorem itself completely guaranties a validity of the EGHK theorem. The original proof of the DGHK theorem can be treated in a similar fashion.  

In order to proceed with the EGHK theorem, let us consider an $N$-electron system defined by the Hamiltonian $H_v^N = T_e^N + V_{ee}^N + V^N$ where $T_e^N$ is the kinetic energy operator of $N$ electrons, $V_{ee}^N$ is the corresponding interelectronic Coulomb operator, 
and $V^N = \Sigma_{i=1}^N v({\bf r}_i)$ is the total external potential. Let $\mid$1$\rangle$ and $\mid$2$ \rangle$ be the ground and the first excited states of $H_v^N$, both nondegenerate, correspondingly described by the normalized wavefunctions $\Psi_1({\bf r}_1, {\bf r}_2, ..., {\bf r}_N) \in {\cal H}^1(\Re^{3N})$ and $\Psi_2({\bf r}_1, {\bf r}_2, ..., {\bf r}_N) \in {\cal H}^1(\Re^{3N})$ (spins are omitted for simplicity; all notations used throughout this Note are defined in Ref. [5]). The $k$-state one-electron density $\rho_k({\bf r})$ $(k = 1, 2)$ is a functional of $v({\bf r})$ [1,2] since $H_v^N$ is explicitly determined by $v({\bf r})$ provided by the given $N, T_e^N$, and $V_{ee}^N$.  

{\bf{Proposal 1}} (EGHK theorem [3]): If there exist two ensemble one-electron densities,
\begin{equation}
\rho({\bf r}) = (1 - w)\rho_1({\bf r}) + w \rho_2({\bf r}),
\label{1}
\end{equation}
corresponding to the Hamiltonian $H_v^N$, and 
\begin{equation}
\rho^\prime({\bf r}) = (1 - w)\rho_1^\prime({\bf r}) + w \rho_2^\prime({\bf r}),
\label{2}
\end{equation}
composed of the weighted one-electron densities $\rho_1^\prime({\bf r}) $ and $\rho_2^\prime({\bf r})$ of the ground state $\mid$$1^\prime\rangle$ and the first excited state $\mid$$2^\prime\rangle$ of the Hamiltonian $H_v^{\prime N} = T_e^N + V_{ee}^N + V^{\prime N}$ where $V^{\prime N} = \Sigma_{i=1}^N v^\prime({\bf r}_i)$ and $w \in [0, 1/2]$, then $\rho({\bf r}) \neq \rho^\prime({\bf r})$ provided that $v({\bf r}) \neq v^\prime({\bf r})$ + constant.\\
\noindent
{\bf{Proof}}: The original proof of the EGHK theorem [3] is based on the method of {\it Reductio ad Absurdum} (or of indirect proof or proof by contradiction; shortly {\it R. A. A.}). Let us first assume the existence of two external potentials $v({\bf r})$ and $v^\prime({\bf r})$ which determine the Hamiltonians $H_v^N$ and $H_v^{\prime N}$ associated with two different $N$-electron systems, such that 
\begin{equation}
v({\bf r}) \neq v^\prime({\bf r}) + \mbox{constant} 
\label{3}
\end{equation}
(premise or proposition ${\mathcal{P}}_1$). It is further assumed (premise ${\mathcal{P}}_2$) that $H_v^N$ and $H_v^{\prime N}$ possess the ground- and first-excited states, $\mid$1$\rangle$ and $\mid$2$\rangle$, and $\mid$$1^\prime\rangle$ and $\mid$$2^\prime\rangle$, respectively. Define the corresponding $N$-electron density matrices with a given $w \in [0, 1/2]$, 
\begin{equation}
D = (1 - w) \mid1\rangle\langle1\mid +\, w \mid2\rangle\langle2\mid
\label{4}
\end{equation}
and 
\begin{equation}
D^\prime = (1 - w) \mid1^\prime\rangle\langle1^\prime\mid +\, w \mid2^\prime\rangle\langle2^\prime\mid.
\label{5}
\end{equation}
(3) implies that [3] \\
\begin{equation}
\mid 1\rangle \neq\; \mid 1^\prime\rangle. 
\label{6}
\end{equation}
To proceed with the {\it R. A. A.} proof of this theorem, the to-be-refuted assumption, i. e., the negation of the desired statement, is chosen as the following premise [3]
\begin{equation}
\rho({\bf r}) = \rho^\prime({\bf r}) \equiv \rho({\bf r}).
\label{7}
\end{equation} 

Applying the Rayleigh-Ritz variational principle (see, e. g., Eq. (4) of Ref. [6]) to Eqs. (4) and (5), one derives a pair of the following inequalities 
\begin{eqnarray}
\mbox{Tr}(D H_v^{N}) &<& \mbox{Tr}(D^\prime H_v^{\prime N}) + \int d^3 {\bf r} [v({\bf r}) - v^\prime({\bf r})] \rho({\bf r}) \nonumber \\
\mbox{Tr}(D^\prime H_v^{N \prime}) &<& \mbox{Tr}(D H_v^{N}) + \int d^3 {\bf r} [v^\prime({\bf r}) - v({\bf r})] \rho({\bf r})
\label{8}
\end{eqnarray}
(Eqs. (7) and (8) of Ref. [3]). Adding them to each other leads to the contradiction,
\begin{equation}
\mbox{Tr}(D H_v^{N}) + \mbox{Tr}(D^\prime H_v^{\prime N}) < \mbox{Tr}(D H_v^{N}) + \mbox{Tr}(D^\prime H_v^{\prime N}),
\label{9}
\end{equation}
as derived from the premises ${\mathcal{P}}_1$ and ${\mathcal{P}}_2$, and the negation (7). Equation (9) is absurd. Its absurdity can be resolved, as Gross et al. conclude [3], by asserting that the to-be-refuted assumption (7) is false. Hence, for a given $w \in [0, 1/2]$, the external potential $v({\bf r})$ is, to within a constant, a unique functional of $\rho({\bf r})$, and since, in turn, $v({\bf r})$ determines $H_v^N$, the ensemble expectation value of $ T_e^N\; +\; V_{ee}^N$ is a functional of the ensemble one-electron density $\rho({\bf r})$. {\em Q. E. D.}

According to the {\it R. A. A.} method [7], the conjunction (hereafter denoted by \&, following Suppes [7a]; usually by $\wedge$) of the premises ${\mathcal{P}}_1$ (see note [8]) and ${\mathcal{P}}_2$ comprises the set {\bf I} of the initial premises given as true. To decide if these premises are consistent [7], deduce from them some formal consequences or implications which are also true:\\
\noindent
{\bf 1.} ${\mathcal{P}}_1\; \&\; {\mathcal{P}}_2 \rightarrow {\mathcal{Q}}_1 \equiv (6)$.\\
\noindent
{\bf 2.} If $v({\bf r})$ and $v^\prime({\bf r})$ are both of Coulomb form (premise ${\mathcal{P}}_3 \in {\bf I}$, by definition), that is, 
\begin{eqnarray}
v({\bf r}) &=& -\Sigma_{\alpha=1}^M \frac{Z_\alpha}{\mid{\bf r} - {\bf R}_\alpha\mid}, \nonumber \\
v^\prime({\bf r}) &=& -\Sigma_{\alpha=1}^{M \prime} \frac{Z_\alpha^\prime}{\mid{\bf r} - {\bf R}_\alpha^\prime\mid} 
\label{10}
\end{eqnarray}
(the $\alpha$th nucleus with the nuclear charge $Z_\alpha$ is placed at ${\bf R}_\alpha \in \Re^3$, and similarly for the primed quantities) then, according to the Kato electron-nuclear cusp theorem [9], from the conjunction ${\mathcal{P}}_1\; \&\; {\mathcal{P}}_2\; \&\; {\mathcal{P}}_3$ one deduces the conditional proposition ${\mathcal{Q}}_2$:
\begin{eqnarray}
\frac{d}{dr_i} \mid k({\bf r}_1, {\bf r}_2, ..., {\bf r}_{i-1}, r_i, {\bf r}_{i+1},..., {\bf r}_N )\rangle^{av_i} \mid_{r_i = R_\alpha} &=& -Z_\alpha \mid k({\bf r}_1, {\bf r}_2, ..., {\bf r}_{i-1}, {\bf R}_\alpha, {\bf r}_{i+1},..., {\bf r}_N )\rangle, \nonumber \\ 
\frac{d}{dr_i} \mid k^\prime({\bf r}_1, {\bf r}_2, ..., {\bf r}_{i-1}, r_i, {\bf r}_{i+1},..., {\bf r}_N ) \rangle^{av_i} \mid_{r_i = R_\alpha^\prime} &=& -Z_\alpha^\prime \mid k^\prime({\bf r}_1, {\bf r}_2, ..., {\bf r}_{i-1}, {\bf R}_\alpha^\prime, {\bf r}_{i+1},..., {\bf r}_N )\rangle, \nonumber \\
\label{11}
\end{eqnarray}
where the superscript `$av_i$' indicates the average of $\mid k \rangle$ or $\mid k^\prime\rangle$ over an infinitesimally small ball centered at ${\bf r}_i$ ($i = 1, 2, ..., N$) and $k, k^\prime = 1,2$.\\
\noindent
{\bf Corollary 1}: ${\mathcal{Q}}_2 \rightarrow {\mathcal{Q}}_1$.\\
\noindent
{\bf 3}. ${\mathcal{P}}_1\; \&\; {\mathcal{P}}_2\; \&\; {\mathcal{P}}_3 \rightarrow {\mathcal{Q}}_3$:
\begin{equation}
\rho({\bf r}) \neq \rho^\prime({\bf r})
\label{12}
\end{equation}
where the ensemble one-electron densities 
$\rho({\bf r})$ and $\rho^\prime({\bf r})$ are defined by Eqs. (1) and (2). \\
\noindent 
{\bf Proof}: ${\mathcal{P}}_3$ implies that the Kato theorem holds for the corresponding one-electron densities:
\begin{eqnarray}
\frac{d}{dr} \rho_k^{av}(r) \mid_{r = R_\alpha} &=& -2Z_\alpha \rho_k({\bf R}_\alpha), \nonumber \\
\frac{d}{dr} \rho_k^{\prime av}(r) \mid_{r = R_\alpha^\prime} &=& -2Z_\alpha^\prime \rho_k^\prime({\bf R}_\alpha^\prime),  
\label{13}
\end{eqnarray}
where the superscript `$av$' indicates the average of $\rho_k$ or $\rho_k^\prime$ over an infinitesimally small ball centered at ${\bf r}$. If, in addition, the premise ${\mathcal{P}}_1$ is an authentic truth, the ensemble one-electron densities $\rho({\bf r})$ and $\rho^\prime({\bf r})$ obey different Kato cusps: 
\begin{eqnarray}
\frac{d}{dr}\rho^{av}(r)\mid_{r = R_\alpha} &=& (1 - w)\frac{d}{dr}\rho_1^{av}(r)\mid_{r = R_\alpha} +\; w\frac{d}{dr}\rho_2^{av}(r)\mid_{r = R_\alpha} \nonumber \\
&=& -2Z_\alpha [(1 - w)\rho_1({\bf R}_\alpha) + w\rho_2({\bf R}_\alpha)] = -2Z_\alpha \rho({\bf R}_\alpha), \nonumber \\
\frac{d}{dr}\rho^{\prime av}(r)\mid_{r = R_\alpha^\prime} &=& (1 - w)\frac{d}{dr}\rho_1^{\prime av}(r)\mid_{r = R_\alpha^\prime} +\; w\frac{d}{dr}\rho_2^{\prime av}(r)\mid_{r = R_\alpha^\prime} \nonumber \\
&=& -2Z_\alpha^\prime [(1 - w)\rho_1^\prime({\bf R}_\alpha^\prime) + w\rho_2^\prime({\bf R}_\alpha^\prime)] = -2Z_\alpha^\prime \rho^\prime({\bf R}_\alpha^\prime).  
\label{14}
\end{eqnarray}
Therefore, they distinguish from each other that implies ${\mathcal{Q}}_3$. {\em Q. E. D.}

According to the {\it R. A. A.} method, the next step of the proof by contradiction, after asserting of the set {\bf I} of the initial premises and verifying their authentical truths, is to introduce ``the negation of the desired conclusion as a new premise" (see, e. g., Ref. [7a], p. 39) and to prove then that this new premise leads to a logical contradiction with the set {\bf I}. The total set {\bf J} of the considered premises is the following:\\
\noindent
{\bf (i)} ${\mathcal{P}}_1\; \&\; {\mathcal{P}}_2\; \&\; {\mathcal{P}}_3$;\\
\noindent
{\bf (ii)} the to-be-refuted premise ${\mathcal{-S}}$ given by Eq. (7).

Within the {\it R. A. A.} method, the proposition ${\mathcal{-S}}$ (i. e., ``not ${\mathcal{S}}" \equiv$ negation of ${\mathcal{S}}$) is equivalent to the proposition that $\cal{S}$ is false [7,10]. 
The proposition ``$\cal{S}$ is false", that is actually ``$\rho({\bf r}) \neq \rho^\prime({\bf r})$", is therefore explicitly equivalent to the conditional proposition ${\mathcal{Q}}_3$ ($\rho({\bf r}) \neq \rho^\prime({\bf r})$, Eq. (12)) which is deduced in {\bf 3.} and which is true if the premise ${\mathcal{P}}_3$ is true. Hence, $\cal{S}$ cannot be false. This straightforwardly implies the intrinsic falsity of the negation ${\mathcal{-S}}$ that exists within the set {\bf J} and hence does not require any other proof ``beyond" {\bf J} to show it [7] (see also note [11]), by analogy with the Rayleigh-Ritz variational principle applied in Ref. [3]. Equivalently, the set {\bf I} of the initial premises given as true is incompatible (``inconsistent", see p. 36{\it f} of Ref. [7a] and notes [12,13]) with the negation ${\mathcal{-S}}$, and thus, the {\it R. A. A.} proof cannot be rigorously mantained in the way suggested in Ref. [3]. This proves the following \\
\noindent
{\bf Proposal 2}: The original proof of the EGHK theorem [3] via the {\it R. A. A.} method is self-contradictory for the class of many-electron systems with Coulomb-type external potentials.

In some sense, {\bf Proposal 2} means that the original proof of the EGHK theorem for Coulomb-type external potentials is superfluous since it starts with an obvious contradiction. It is however shown above that ${\mathcal{P}}_1\; \&\; {\mathcal{P}}_2\; \&\; {\mathcal{P}}_3 \rightarrow {\mathcal{Q}}_3$, and therefore, one logically derives\\
\noindent
{\bf Corollary 2}: If the Kato theorem holds, for the class of many-electron systems with Coulomb-type external potentials the EGHK statement is correct.       

In conclusion, the present Note demonstrates that the Kato theorem itself provides the direct proof of the EGHK theorem for $N$-electron systems with the Coulomb class of external potentials. A similar reasoning can be trivially applied to the original proof of the DGHK theorem given in Ref. [4] to show that its original proof is self-contradictory too, as though the statement of the DGHK theorem is correct, due to the Kato theorem. With regard of the EGHK theorem, it is also worth mentioning the super-Hamiltonian approach to the ensemble-density functional theory that was originally introduced by Theophilou [14] and further developed by Katriel [15], and also in Ref. [16] within the local-scaling-transformation method of the density functional theory [2c, 17].
\vspace{1cm}

\centerline{\bf {Acknowledgment}}
\vspace{0.25cm}

I gratefully thank all colleagues for the helpful and inspired discussions of my early work [5]. I also thank Francoise Remacle for kind hospitality and F.R.F.C. 2.4562.03F (Belgium) for fellowship. 


\vspace{0.5cm}

\begin{thebibliography}{99}
\bibitem{a1} P. Hohenberg and W. Kohn, Phys. Rev. {\bf 136}, B864 (1964). 
\bibitem{a2} (a) R. G. Parr and W. Yang, {\em Density-Functional Theory of Atoms and Molecules} (Oxford University Press, New York, 1989). (b) R. M. Dreizler and E. K. U. Gross, {\em Density Functional Theory} (Springer-Verlag, Berlin, 1990). (c) E. S. Kryachko and E. V. Lude\~na, {\em Energy Density Functional Theory of Many-Electron Systems} (Kluwer, Dordrecht, 1990). (d) N. H. March, {\em Electron Density Theory of Atoms and Molecules} (Academic, London, 1992). 
\bibitem{a3} E. K. U. Gross, L. N. Oliviera, and W. Kohn, Phys. Rev. A {\bf 37}, 2809 (1988). 
\bibitem{a4} R. van Leeuwen, Adv. Quantum Chem. {\bf 43}, 24 (2003). 
\bibitem{a5} E. S. Kryachko, Int. J. Quantum Chem. {\bf 103}, 818 (2005), quant-ph/0504114.
\bibitem{a6} E. K. U. Gross, L. N. Oliviera, and W. Kohn, Phys. Rev. A {\bf 37}, 2805 (1988).
\bibitem{a7} (a) P. Suppes, {\em Introduction to Logic} (Van Nostrand, Princeton, 1967). (b) A. N. Whitehead and B. Russell, {\em Principia Mathematica} (Cambridge University Press, Cambridge, 1962). (c) I. M. Copi {\em Symbolic Logic} (Macmillan, New York, 1965). (d) A. N. Prior, {\em Formal Logic} (Clarendon, Oxford, 1962). (e) D. Scherer, Mind {\bf LXXX}, 247 (1971). (f) J. M. Lee, Notre Dame J. Formal Logic {\bf XIV}, 381 (1973). (g) L. C. D. Kulathungam, Notre Dame J. Formal Logic {\bf XVI}, 245 (1975).   
\bibitem{a8} Rigorously speaking, a constant term on the rhs of (3) should be eliminated as contradicting to the vanishing asymptotic behavior of the external potential of an arbitrary (unconfined) many-electron system at infinity. 
\bibitem{a9} T. Kato, Commun. Pure Appl. Math. {\bf 10}, 151 (1957).
\bibitem{a10} By definition, ``not ${\mathcal{S}}"$ is ``the proposition that is true when ${\mathcal{S}}$ is false and false when ${\mathcal{S}}$ is true." See, e. g., G. E. M. Anscombe, {\em An Introduction to Wittgenstein's Tractatus} (Hutchinson University Library, London, 1959). 
\bibitem{a11} Two propositions ``are said to be {\it contradictory} if one is the negation of the other." (p. 37 of Ref. [7a]).
\bibitem{a12} G\"odel [K. G\"odel, {\em Obras Completas} (Alianza Editorial, Madrid, 1981)] defines that a set of propositions is consistent if it cannot simultaneously deduce a proposition and its negation.  
\bibitem{a13} About contradictions in physics see Y. Aharonov and D. Rohrlich, {\em Quantum Paradoxes. Quantum Theory for the Perplexed} (Wiley-VCH, Weinheim, 2005). 
\bibitem{a14} (a) A. K. Theophilou, J. Phys. C: Solid State {\bf 12}, 5419 (1979). (b) V. N. Glushkov and A. K. Theophilou, Phys. Rev. A {\bf 64}, 064501 (2001).  
\bibitem{a15} (a) J. Katriel, J. Phys. C: Solid State {\bf 13}, L375 (1980). (b) J. Katriel and F. Zahariev, Phys. Rev. A {\bf 65}, 024501 (2002).
\bibitem{a16} (a) E. S. Kryachko, E. V. Lude\~na, and T. Koga, J. Math. Chem. {\bf 11}, 325 (1992). (b) T. Koga, J. Chem. Phys. {\bf 95}, 4306 (1991). (c) E. S. Kryachko and  E. V. Lude\~na, J. Mol. Struct. (Theochem) {\bf 287}, 1 (1993). (d) T. Koga, J. Math. Chem. {\bf 14}, 207 (1993).
\bibitem{a17} (a) E. V. Lude\~na, R. Lopez-Boada, J. Maldonado, E. Valderrama, E. S. Kryachko, T. Koga, and J. Hinze, Int. J. Quantum Chem. {\bf 56}, 285 (1995). (b) E. S. Kryachko, in {\em New Methods in Quantum Theory}, edited by C. A. Tsipis, V. S. Popov, D. R. Herschbach, and J. S. Avery (NATO ASI Series, Kluwer, Dordrecht, 1995), p. 339.
\end{thebibliography}
\end{document}